\documentclass[prl,twocolumn,showpacs,groupedaddress]{revtex4}
\usepackage{graphicx}
\usepackage{epsfig,color}
\usepackage{dcolumn}

\begin{document}

\title{Electron-phonon interaction in the lamellar cobaltate Na$_x$CoO$_2$}

\author {A.~Donkov$^{1}$}
\author {M.M.~Korshunov$^{1,2}$}
\author {I.~Eremin$^{1,3}$}
 \affiliation {$^1$Max-Planck-Institut f\"{u}r Physik komplexer Systeme, D-01187 Dresden, Germany}
 \affiliation {$^2$L.V. Kirensky Institute of Physics, Siberian Branch of Russian Academy of Sciences, 660036 Krasnoyarsk, Russia}
 \affiliation {$^3$Institute f\"{u}r Mathematische und Theoretische Physik, TU Braunschweig, D-38106 Braunschweig, Germany}

\author {P.~Lemmens$^{1}$}
\author {V.~Gnezdilov$^{2}$}
\author {F.C.~Chou$^{3}$}
\author {C.T.~Lin$^{4}$}
 \affiliation {$^1$Institute for Physics of Condensed Matter, TU Braunschweig, D-38106 Braunschweig, Germany}
 \affiliation {$^2$B. I. Verkin Inst. for Low Temperature Physics, NASU, 61164 Kharkov, Ukraine}
 \affiliation {$^3$ Center for Condensed Matter Sciences, National Taiwan University, 
 Taipei 10617, Taiwan}
 \affiliation {$^4$Max Planck Institute for Solid State Research, D-70569 Stuttgart, Germany}

\date{\today}

\begin{abstract}
We study theoretically and experimentally the dependence of the electron-phonon
interaction in Na$_x$CoO$_2$ on the sodium concentration $x$. For the two
oxygen phonon modes found in Raman experiments, $A_{1g}$ and $E_{1g}$, we
calculate the matrix elements of the electron-phonon interaction. Analyzing the
feedback effect of the conduction electrons on the phonon frequency
we compare the calculated and experimentally observed doping dependence of the
$A_{1g}$ mode. Furthermore, due to the momentum dependence of the
electron-phonon coupling for the $E_{1g}$ symmetry we find no renormalization
of the corresponding phonon frequency which agrees with experiment. Our results
shed light on the possible importance of the electron-phonon interaction in
this system.
\end{abstract}


\maketitle

{\it Introduction.} The origin of the unconventional superconductivity in
low-dimensional perovskite systems attracts much attention and belongs to the most
challenging questions of condensed matter physics. The best known example among these
materials are high-$T_c$ cuprate superconductors. There, one of the possible scenarios
for the Cooper-pair formation is the so-called spin fluctuation mechanism. At the same
time, due to the complexity of the transition metal oxides, other energy scales are
present, and their role in the formation of superconductivity remains under debate. This,
in particular, concerns the electron-phonon interaction. For example its relevance for
superconductivity in layered cuprates and the anomalous normal state has been discussed
in, e.g., Ref. \onlinecite{shen}. Despite some progress, a complete understanding of the
physics of electron-phonon coupling in perovskites is still lacking because of the
crystallographic complexities of these materials.

The discovery of superconductivity with T$_c$=4.6K in water intercalated sodium
cobaltate, Na$_x$CoO$_2 \cdot y$H$_2$O \cite{Takada2003}, is of great interest on its
own and also because of similarities with layered cuprates. The sodium cobaltate has a
quasi-two-dimensional layered structure with CoO$_2$ layers and rich phase diagram as a
function of the Na concentration, which includes superconductivity at $x \approx 0.3$,
an insulating phase at $x\sim 0.5$, and unusual magnetism for $x\geq 0.6$
\cite{Foo2004}. There is also increasing experimental and theoretical evidence for
unexpected strong correlation effects as the cobaltates approach the band insulting
limit at $x=1$ \cite{bernh07,lee,merino,haerter,Korshunov2007,daghofer,marianetti,wang}.
In Na$_x$CoO$_2$ the Na ions reside between the CoO$_2$ layers, with Co ions forming a
triangular lattice, and donate $x$ electrons to the partially filled Co-$d(t_{2g})$
orbitals. Due to the presence of a trigonal crystalline electric field, the $t_{2g}$
levels split into the higher lying $a_{1g}$ singlet and the two lower lying $e'_g$
states \cite{djs2000}. Angle-Resolved Photo-Emission Spectroscopy (ARPES) \cite{mzh2004,
hby2004} reveals a doping dependent evolution of the Fermi surface, which shows no sign
of the $e'_g$ hole pockets for $0.3 \le x \le 0.8$. The observed Fermi surface is
centered around the $\Gamma$ point and has mostly $a_{1g}$ character. It has been argued
that such an effect may arise due to strong electronic correlations
\cite{Korshunov2007,Zhou2005}, or Na induced disorder \cite{Singh2006}, however, no
consensus in the literature has been reached yet (see, for example, \cite{Ishida2005,Balicas2006, Chou07}).

Despite of intensive studies of the electronic and magnetic properties little is known
about the phonon excitations and their doping evolution in Na$_x$CoO$_2$. At the same
time, due to the relatively low superconducting transition temperature the possible
relevance of phonons for superconductivity cannot be neglected. For example, the role of
the electron-phonon coupling in Na$_x$CoO$_2$ has been discussed in the context of its
relevance to superconductivity and charge ordering on the basis of a $t-V$ model
\cite{greco01}. In addition, due to some similarity with high-$T_c$ cuprates the
understanding of the phonon renormalization in the sodium cobaltates is of great
importance. Initially, the effect of renormalization of the optical phonons by the
conduction electrons in layered cuprates has been considered in Ref.
\onlinecite{zwicknagl}.

In this Rapid Communication we investigate the electron-phonon interaction in the
Na$_x$CoO$_2$ as a function of doping concentration and its superconducting relative by
means of Raman spectroscopy. We observe two oxygen phonon modes at small wave vectors
with $A_{1g}$ and $E_{1g}$ symmetries. Then we derive the diagonal and off-diagonal
electron-phonon matrix elements for these modes. Calculating the renormalization of the
phonon frequencies by conduction electrons we compare our results with the doping
dependent evolution of the $A_{1g}$ mode and obtain the electron-phonon coupling
constant $g^{A_{1g}}_{off} = 3$meV. Due to the structure of the electron-phonon matrix
element for the $E_{1g}$ mode we obtain no doping dependence of the corresponding phonon
frequency in good agreement with experiment. Our results shed light on the possible role
of the electron-phonon interaction in this compound.

\begin{figure}
\includegraphics[angle=0,width=1.0\linewidth]{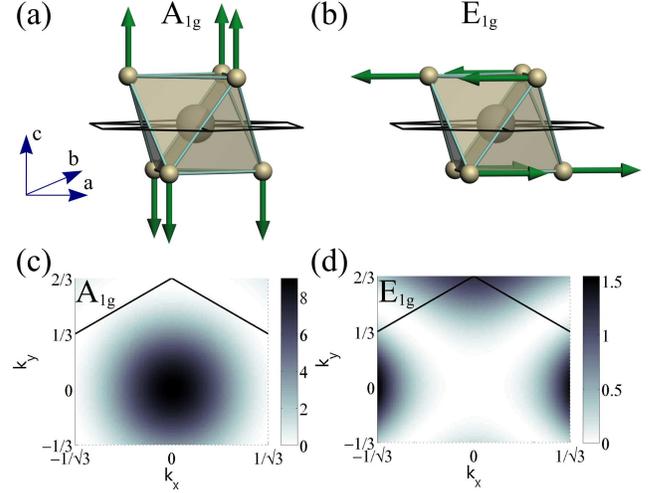}
\caption{(Color online) Schematic illustration of the $A_{1g}$ (a)
and $E_{1g}$ (b) phonon modes. The arrows indicate the oxygen
displacement in the CoO$_6$ octahedra. On the left side of (a) we indicate the
crystallographic $a$, $b$, and $c$ directions.
(c)-(d) The calculated momentum dependence of the electron-phonon
structure factors, $\left[F_{\bf q}^\Gamma\right]^2$, in the first BZ for the
$A_{1g}$ and $E_{1g}$ phonon modes, respectively.} \label{fig:modes}
\end{figure}

\begin{figure}
\includegraphics[angle=0,width=0.78\linewidth]{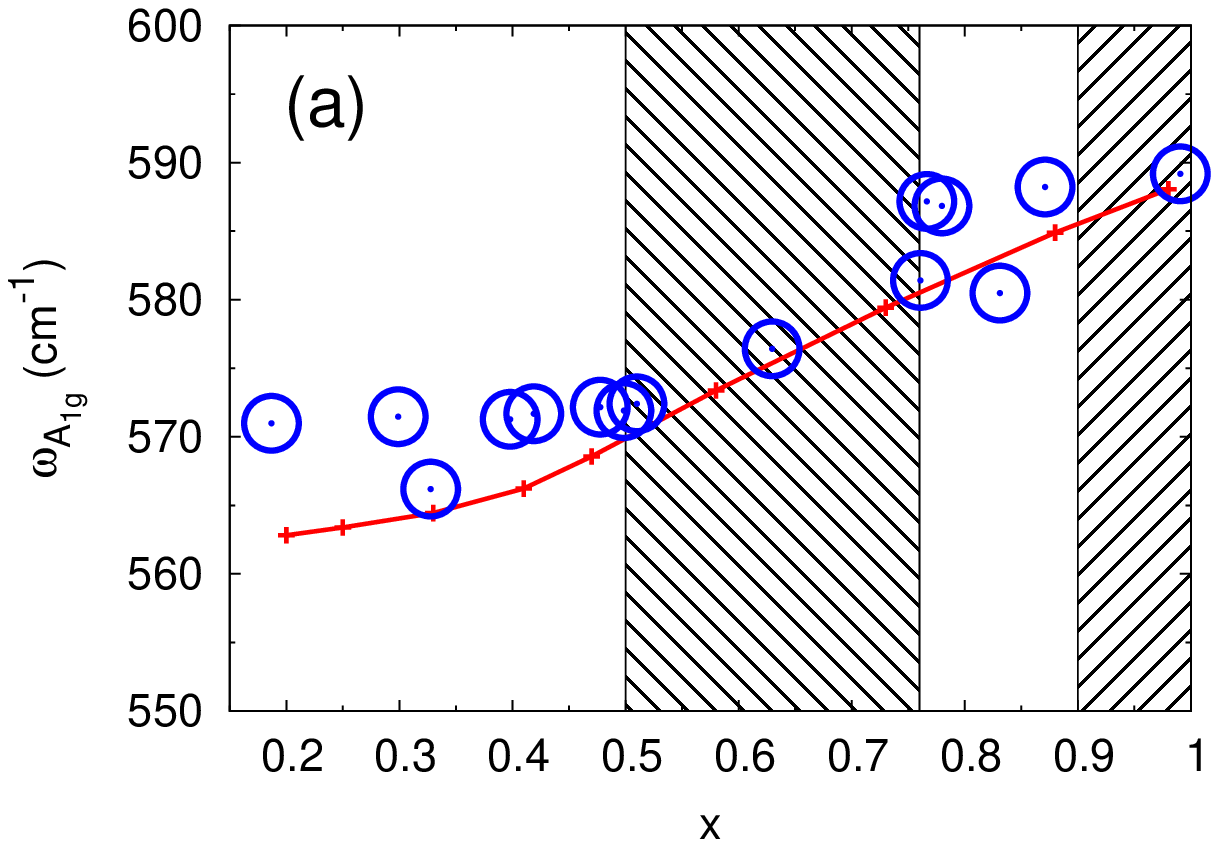}
\includegraphics[angle=0,width=0.78\linewidth]{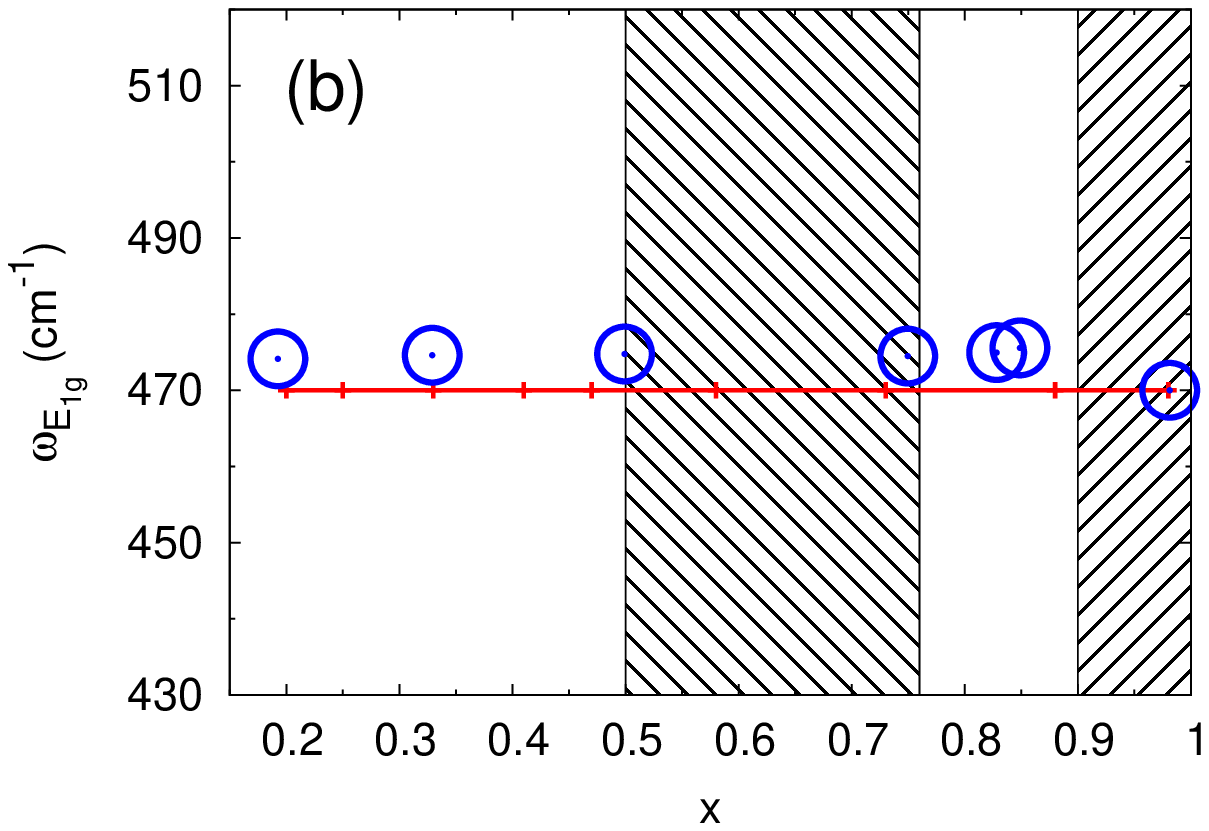}
\caption{(Color online) The experimental Raman data at 10K is shown by
(blue) circles. The calculated doping dependence
of the $A_{1g}$ (a) and $E_{1g}$ (b) phonon modes is shown by red crosses, the solid curve is a guide to the eye. The slight
scattering of the experimental points around $x=0.5$, $x=0.76$ and $x=0.9$ (shown by the
dashed areas) are the result of the different crystallographic phases around these
points\cite{Lin2006}. These structural modifications only weakly affect the
electron-phonon coupling as it mainly involves Na ordering. Measurements of $A_{1g}$ mode at 290K 
(not shown) follow a similar trend as the 10K data.} \label{fig:ph_renorm}
\end{figure}

{\it Experiment details.} Raman scattering experiments have been performed in
quasi-backscattering geometry on freshly cleaved single crystal surfaces. The sample
have been fully characterized using basic thermodynamic as well as spectroscopic
techniques \cite{Lemmens2004,Lemmens2006,Bayrakci2004,Chen2004,Lin2005,Lin2006}. After
cleavage the crystals were rapidly cooled down in He exchange gas to prevent
degradation. In Na$_x$CoO$_2$ in-plane $E_{1g}$ and out-of-plane $A_{1g}$ oxygen modes
have been observed in Raman scattering \cite{Iliev2004,Lemmens2006,Lemmens2007} and the
corresponding oxygen displacements are depicted in Fig.~\ref{fig:modes}(a) and (b). The
Co site is not Raman-active. Modes of the Na sites have not been identified
unambiguously\cite{Iliev2004}. This is probably related to disorder on the partially
filled sites. The two-dimensionality with respect to structure and bonding leads to a
decoupling of the Na and the CoO$_2$ layers. The observed doping-dependence of the
$A_{1g}$ and $E_{1g}$ oxygen phonon frequencies are shown in Fig.~\ref{fig:ph_renorm}.
The cross-over from one to the other crystallographic phases (shaded areas) given by a
different occupation of the Na sites leads for the $A_{1g}$ modes to small additional
frequency shifts and for the $E_{1g}$ modes to new modes with a larger energy off-set.
The latter are omitted for clarity. The two phonon modes display a markedly different
doping dependence.

{\it Tight-binding model.} To describe the electronic subsystem we use a
tight-binding $t_{2g}$-band model with parameters (in-plane hoppings and the
single-electron energies) derived previously from the {\it ab-initio} LDA
(Local Density Approximation) calculations using projection procedure for
$x=0.33$ \cite{Korshunov2007}.

The free-electron Hamiltonian of the $t_{2g}$-band model in a hole
representation is given by
\begin{equation}
H_0 = - \sum\limits_{{\bf k},\alpha ,\sigma } {\left( {\epsilon
^\alpha - \mu } \right)n_{{\bf k} \alpha \sigma } } -
\sum\limits_{{\bf k}, \sigma} \sum\limits_{\alpha, \beta}
t_{{\bf k}}^{\alpha \beta } d_{{\bf k} \alpha \sigma }^\dag d_{{\bf k} \beta \sigma},
\label{eq:H0}
\end{equation}
where $n_{{\bf k} \alpha \sigma} = d_{{\bf k} \alpha \sigma}^\dag d_{{\bf k}
\alpha \sigma}$, $d_{{\bf k} \alpha \sigma}$ ($d_{{\bf k} \alpha \sigma}^\dag$)
is the annihilation (creation) operator for the $t_{2g}$-hole with spin
$\sigma$, orbital index $\alpha$, and momentum ${\bf k}$, $t_{{\bf k}}^{\alpha
\beta}$ is the hopping matrix element, and $\epsilon^{\alpha}$ is the
single-electron energy. To obtain the dispersion we diagonalize the Hamiltonian
(\ref{eq:H0}) calculating the chemical potential $\mu$ self-consistently. Due
to the non-zero inter-orbital hopping matrix elements, $a_{1g}$ and $e'_{g}$
bands are hybridized. However, only one of the hybridized bands crosses the
Fermi level. We refer to the diagonalized bands as $\varepsilon^{\alpha'}_{\bf
k}$ with the new orbital index $\alpha'$.

{\it Electron-phonon interaction.} In analogy to previous considerations for
cuprates \cite{Nazarenko1996,Schneider2005}, we derive the electron-phonon
matrix elements, $g_{\bf kq}$, for the $A_{1g}$ and $E_{1g}$ phonon modes
depicted in Fig.~\ref{fig:modes}(a) and (b). Namely, to obtain the main
contribution to the diagonal (intraband) part of the electron-phonon
interaction we expand the Coulomb energy between Co and oxygen , $H_C = \frac{e
e^*}{\epsilon} \sum_{i, \alpha', \sigma, \gamma} c^\dag_{i \alpha' \sigma} c_{i
\alpha' \sigma} \left( \frac{1}{|{\bf R}_i-{\bf r}_{i,\gamma}|} +
\frac{1}{|{\bf R}_i-{\bf r}_{i,-\gamma}|} \right)$, in the small displacements
of the oxygen ions. Here, $e$ is the electron charge, $e^*=-2e$ is the oxygen
ion charge, $\epsilon$ is the dielectric constant, ${\bf R}_i$ are the Co ion
positions, ${\bf r}_{i,\gamma}$ are the vector positions of the vibrating
oxygens, and index $\gamma$ ($-\gamma$) labels the three oxygen positions
within CoO$_6$ unit cell above (below) the Co layer. Here, $c^{\dag}_{i \alpha'
\sigma}$ refers to the diagonal form of the Hamiltonian (\ref{eq:H0}).
After introducing the creation (annihilation) operator $b^{\dag}_{\bf -q}$
($b_{\bf q}$) for the phonon with momentum ${\bf q}$, we arrive to the
following form of the electron-phonon interaction
\begin{equation}
H^{diag}_{el-ph} = \sum_{{\bf k}, {\bf q}, \alpha', \sigma} g^{\alpha'
\alpha'}_{\bf q} c^\dag_{{\bf k} \alpha' \sigma} c_{{\bf k-q} \alpha' \sigma} (b_{\bf q}+b^\dag_{\bf -q}).
\end{equation}
For the sake of simplicity we assume the diagonal electron-phonon
interaction is independent on the orbital index $\alpha'$. Thus, for
the $A_{1g}$ and $E_{1g}$ optical Raman-active phonon oxygen modes
one finds $g^{A_{1g}}_{\bf q} = g^{A_{1g}}_{diag} F^{A_{1g}}_{\bf
q}$, $g^{E_{1g}}_{\bf q} = g^{E_{1g}}_{diag} F^{E_{1g}}_{\bf q}$,
where the structure factors of the electron-phonon interaction are
\begin{eqnarray}
F^{A_{1g}}_{\bf q} & = & \cos \frac{q_1-q_2}{3} + \cos \frac{q_1+q_3}{3} + \cos \frac{q_2+q_3}{3}, \\
F^{E_{1g}}_{\bf q} & = & \cos \frac{q_1-q_2}{3} - \frac{1}{2} \left[
\cos \frac{q_1+q_3}{3} + \cos \frac{q_2+q_3}{3} \right].
\end{eqnarray}
Here, $q_1 = (\sqrt{3}/2) q_x - \frac{1}{2} q_y$,  $q_2 = q_y$, $q_3
= (\sqrt{3}/2) q_x + \frac{1}{2} q_y$, in units of $2\pi/a$ with $a$
being the in-plane lattice constant, $g^\Gamma_{diag} = - \frac{e
e^*}{\epsilon} \frac{2L_\Gamma}{\sqrt{d^2 +l^2}^3}
\sqrt{\frac{\hbar}{2 M \omega_\Gamma}}$, where $\omega_\Gamma$ is
the corresponding bare phonon frequency ($\Gamma=A_{1g}, E_{1g}$),
$L_{A_{1g}}=d=a / \sqrt{6}$ is the distance between the Co and the
oxygen plane, $L_{E_{1g}}=l=a / \sqrt{3}$ is the planar distance
between Co and oxygen, and $M$ is the oxygen mass. Assuming that in
the band insulator, Na$_{x=1}$CoO$_2$, the renormalization of the
phonons by the conduction electrons is absent, we use
$\omega_{A_{1g}}=589 cm^{-1}$ and $\omega_{E_{1g}}=470 cm^{-1}$.
These values are close to those obtained by the first principles
calculations \cite{Li}. The resulting momentum dependence of the
structure factors for the both modes is shown in
Fig.~\ref{fig:modes}(c) and (d). Interestingly, one sees that while
the $g_{\bf q}$ for the $A_{1g}$ mode shows a maximum at the BZ
center, the corresponding $g_{\bf q}$ for the $E_{1g}$ mode vanishes
there. Therefore, for ${\bf q}=0$ the electron-phonon coupling for
the $E_{1g}$ channel is zero. Taking $\epsilon \sim 20$ we estimate
$g^{A_{1g}}_{diag} \approx 0.05$eV.

The off-diagonal (interband) contribution to the electron-phonon interaction
arises mainly from the modulation of the inter-orbital Co-Co hopping matrix
element via oxygen. Assuming the linear terms in the expansion of the nearest
neighbors hopping matrix element $t_{ij}^{\alpha \beta}(u_{\gamma}) =
t_{ij}^{\alpha \beta}+ V^{\alpha \beta} u_{\gamma}$ over the oxygen
displacements $u_{\gamma}$, one obtains:
\begin{equation}
H^{off}_{el-ph} = \sum_{{\bf k}, {\bf q}, \alpha' \neq \beta', \sigma}
g^{\alpha' \beta'}_{\bf kq} c^\dag_{{\bf k}\alpha'\sigma } c_{{\bf
k-q}\beta'\sigma } (b_{\bf q}+b^\dag_{\bf -q}).
\end{equation}
where $g^{\alpha' \beta'}_{\bf kq} = g^\Gamma_{off} F^\Gamma_{\bf q}
(\gamma({\bf k})+\gamma({\bf k+q}))$ with $\gamma({\bf k}) = \cos
k_2 + \cos k_3 + \cos k_1$ being the Co lattice structure factor.
Again one could see that for ${\bf q}=0$ the off-diagonal
electron-phonon coupling for the $E_{1g}$ channel is zero due to the
momentum dependence of the electron-phonon structure factor,
$F^{E_{1g}}_{\bf q}$. Therefore, in Raman experiments which probes
${\bf q}=0$ response this mode shows no doping dependence due to the
coupling to the electronic subsystem. This is also confirmed by the
fact that the observed phonon mode energy for all doping levels lies
close to the value obtained by {\it ab-initio} calculations
\cite{Li}. The only Raman-active optical phonon mode which will
couple to the conduction electrons at ${\bf q}=0$ is the $A_{1g}$
mode.

In the following we consider the renormalization of the $A_{1g}$ phonon. The
corresponding Dyson equation reads:
\begin{equation}
D^{-1}({\bf q},\omega)=D^{-1}_0(\omega)- \Pi({\bf q},\omega),
\label{eq:dyson}
\end{equation}
where $D_0 (\omega) = \frac{2\omega_\Gamma}{\omega^2-\omega_\Gamma^{2} +
i\delta}$ is the momentum-independent bare phonon propagator. The polarization
operator is given by:
\begin{equation}
\Pi({\bf q}, \omega) = - 2 \sum_{\alpha',\beta'} \sum_{\bf k} \left(
g_{\bf kq}^{\alpha' \beta'}\right)^{2}
\frac{f(\varepsilon^{\alpha'}_{\bf k+q} ) -
f(\varepsilon^{\beta'}_{\bf k})}{ \omega -
\varepsilon^{\alpha'}_{\bf k+q} + \varepsilon^{\beta'}_{\bf
k}+i\delta}, \label{eq:Pi}
\end{equation}
where $f(\epsilon)$ is the Fermi function. To find the
renormalization of the bare phonon frequency and to compare the
results to the Raman experiments we set ${\bf q} \to 0$ limit and
solve Eq.~(\ref{eq:dyson}) numerically as a function of the doping
concentration. The main contribution to the renormalization of the
optical phonon modes comes from the interband transitions, {\it
i.e.} terms with $g_{\bf kq}^{\alpha' \neq \beta'}$ while intraband
transitions renormalize the acoustic phonon modes. The results of
our numerical calculations are shown in Fig.~\ref{fig:ph_renorm}.
The doping evolution has been deduced by calculating $\Pi({\bf q}
\to 0, \omega)$ for various $x$ values. We obtain the value of the
off-diagonal electron-phonon interaction $g^{A_{1g}}_{off} \approx
3$meV by comparison to the experimental data points. This value is
an order of magnitude smaller than the diagonal contribution to the
electron-phonon interaction. One sees that the phonon
renormalization changes smoothly as a function of doping. The
renormalization effects tend to vanish close to the band insulator
regime at $x=1$ because all of the Fermi functions entering
Eq.(\ref{eq:Pi}) are equal one and Re$\Pi({\bf q},0) = 0$. Away from
this point Re$\Pi$ will be simply determined by the number of holes
in the system. This explains the linear doping dependence of the
renormalization effects. We also note that the electronic
correlation effects seem to play a minor role for the
renormalization of the Raman-active optical phonon modes. The
situation may differ, however, for the acoustic phonons where the
density of the states at the Fermi level plays the most important
role \cite{dolgov}.

Of course, our estimated value for the electron-phonon interaction
corresponds to the interband scattering which is not directly
related to the superconductivity. Therefore, we use our estimated
value for the diagonal part of the electron-phonon interaction and
obtain using the BCS formula $k_B T_c = 1.14 \omega_{A_{1g}}
\exp\left(-1/N(E_F)g^{A_{1g}}\right)$  T$_c \sim 7$ K for the
$N(E_F) \approx 4.0$ states/eV \cite{djs2000}. This estimate is in agreement with the observed $T_c$ in water intercalated Na$_x$CoO$_2$, which points towards potential relevance of the electron-phonon interaction for the superconductivity in this compound. At the same time, the smooth evolution of the phonon frequencies as a function of the doping concentration on the one hand, and the rich phase diagram of the lamellar cobaltates, on the
other, requires further understanding the role of the phonons in the formation of superconductivity.
One interesting possibility to explore is the possible enhancement of $T_c$ due to resonance levels the may be introduced by the water molecules \cite{Little:1964PR,Ginzburg64,Ginzburg}.
Very recently the so called Suhl-Kondo resonance was suggested as an origin of the Cooper paring in the Na$_x$CoO$_2 \cdot y$H$_2$O \cite{YadaKontani}. In any case a study of the isotope effect on $T_c$ is highly desirable.

\begin{acknowledgments}
We would like to thank P. Fulde, B.~Keimer, I. Mazin, K. Morawetz, E. Schneider, 
A. Yaresko, and G. Zwicknagl for useful discussions. We thank P.
Scheib and A. Boothroyd for contributions to our experimental study.
M.M.K. acknowledges support from INTAS (YS Grant 05-109-4891) and
RFBR (Grants 06-02-16100, 06-02-90537-BNTS). I.E. acknowledges
support from Volkswagen Foundation. The experimental studies have
been supported by DFG and ESF-HFM.
\end{acknowledgments}

\end{document}